\documentclass[
twocolumn,
amsmath,amssymb,
aps,
]{revtex4-2}
\usepackage{graphicx}
\usepackage{hyperref}
\usepackage{dcolumn}
\usepackage{bm}
\usepackage{epstopdf}
\usepackage{romannum}
\usepackage[dvipsnames]{xcolor}

\begin{document}

\preprint{Sun/depairing Jc}

\title{Depairing critical current density and the vortex-free state in FeSe nanobridges}

\author{Yue Sun$^1$}
\email{sunyue@seu.edu.cn}
\author{Yuling Xiang$^1$, Zhixiang Shi$^1$, Tsuyoshi Tamegai$^2$}
\author{Haruhisa Kitano$^3$}
\email{hkitano@phys.aoyama.ac.jp}

\affiliation{%
$^1$School of Physics, Southeast University, Nanjing 211189, China\\
$^2$Department of Applied Physics, The University of Tokyo, Tokyo 113-8656, Japan\\
$^3$Department of Physics and Mathematics, Aoyama Gakuin University, Sagamihara 252-5258, Japan}

\date{\today}

\begin{abstract}
The depairing limit and the vortex-free state in a superconductor is crucial for both the study of supercurrent related physics and the application eliminating noise linked to vortex motion. In this work, we report the evidence of depairing limit and the vortex-free state achieved by geometric constraint in FeSe superconductors. A series of narrow bridges with varying widths at the same location of a single crystal were prepared by the \textquotedblleft pickup\textquotedblright method using successive focused ion beam millings. By simply reducing the width of bridge, the magnitude of critical current density ($J_{\rm{c}}$) is enhanced more than one order, evidence the achievement of depairing limit. Moreover, in the bridge with a width smaller than the penetration depth ($\lambda$), $J_{\rm{c}}$ is found to be robust against magnetic field up to 1 kOe. The field-robust $J_{\rm{c}}$ is a strong piece of evidence for vortex-free state, which is created by the enhancement of lower critical fields due to geometric constraint.          

\end{abstract}

\maketitle
\section{introduction}

The depairing critical current density ($J_{\rm{c}}$) determined by the depairing process of Cooper pairs \cite{Tinkham}, is crucial for both the study of superconducting mechanism and practical application. It directly provides information on the critical velocity of superfluids related to the superconducting (SC) gap magnitude, and the upper limit of the applied current in the application. The depairing process occurs when the kinetic energy of the supercurrent exceeds the condensation energy ($\propto$ SC gap) \cite{Tinkham, Dew-HughesJcReview}. However, it is difficult to be achieved, because the supercurrent tends to pile up at the surfaces and/or edges of the superconductor due to the Meissner effect (Magnetic flux density is the largest at the surface/edge part as the flux lines circle the sample) \cite{Tinkham}. Then, the vortices can enter the superconductor from the surface- and/or edge-part because the current density there is already above the critical value, although the value of $J_{\rm{c}}$ for the whole sample is still much smaller than the depairing $J_{\rm{c}}$ \cite{SkocpolPhysRevB.14.1045,Tinkham}. After the entrance of vortices, the critical current density is determined by the vortex flow, which occurs when the Lorentz force exceeds the pinning force of vortices, and is usually called depinning $J_{\rm{c}}$ \cite{Dew-HughesJcReview,Blatterreview}. 

The entrance of vortices due to the nonuniform supercurrent could be overcome by the geometric constraint. Theoretically, the depairing current density can be obtained when the transverse dimension of the bridge is made small compared to both the coherence length and the penetration depth \cite{Tinkham}. In high-$T_{\rm{c}}$ superconductors, the condition is hard to be fulfilled since their coherence length is too small, typically several nanometers \cite{TsueiRevModPhys.72.969}. In practice, the depairing limit can be achieved if the current is homogeneously distributed in the bridge \cite{Tinkham,NawazPhysRevLettYBCOJdp}, which can be fulfilled when the width of the bridge ($W$) is reduced to the Pearl length $\varLambda$ = 2$\lambda^2/h$, where $\lambda$ is the penetration depth and $h$ is the thickness of the bridge \cite{ClemPearllength}. In this case, the current density is uniformly distributed in the bridge, and the depairing limit can be achieved. Such a method has been successfully applied to achieve the depairing limit in the YBa$_2$Cu$_3$O$_{7-\delta}$, Ba$_{0.5}$K$_{0.5}$Fe$_2$As$_2$, and the $c$-axis bridge of Fe$_{1+y}$Te$_{1-x}$Se$_{x}$ \cite{NawazPhysRevLettYBCOJdp,LiJunAPLdeparingJcBaK122,SundepairingJc} and P-doped BaFe$_2$As$_2$ \cite{MizukoshiPhysRevB.110.104501}. Furthermore, the geometric constraint in a superconductor will also affect the penetration of vortices under magnetic field, and even expel all the vortices to make a vortex-free state. Theoretically, vortices could be expelled from a superconducting thin layer when $W$ is reduced smaller than $\lambda$ due to the enhancement of the lower critical field $\mu_0H_{\rm{c1}}$ \cite{Tinkham}. Experimentally, the beginning of vortex penetration was reported to be enhanced up to $\sim$ 1 T for YBa$_2$Cu$_3$O$_{7-d}$ nanowires with 80 nm wide and 150 nm thick \cite{Victornanolett}. 

The achievement of depairing limit and the vortex-free state is crucial for the study of supercurrent and related physics in superconductivity. It also has potential for applications of the superconducting structures to eliminate noise linked to vortex motion, especially under high fields. It has already been well studied in low-temperature and high-temperature cuprate superconductors \cite{Blatterreview,KuitPhysRevB.77.134504}. However, for another high-temperature superconductors, iron-based superconductors (IBSs), the depairing limit and the vortex-free state remain poorly understood, despite significant progress in application-oriented research. \cite{Hosono2015}.

\begin{figure*}\center
	\includegraphics[width=16cm]{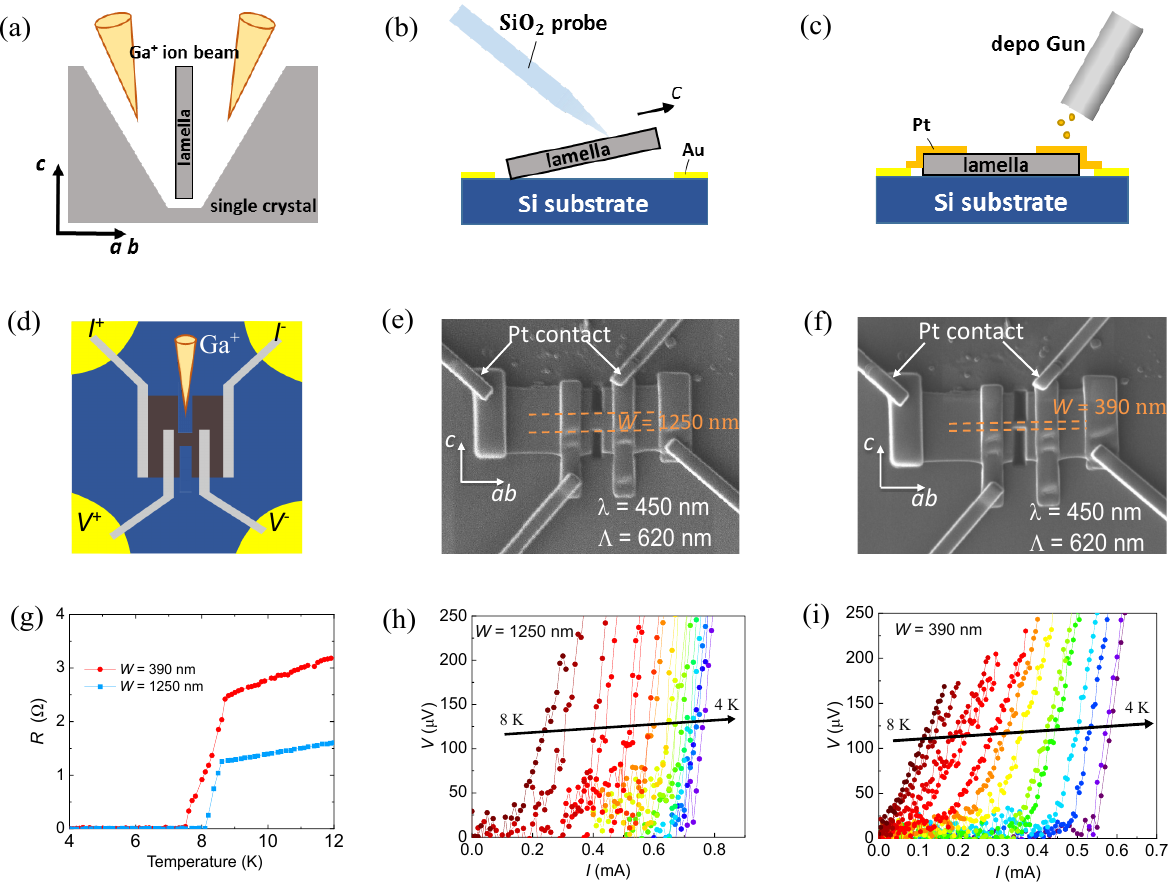}
	\caption{(a)-(d)Schematics of the "pickup" method by using FIB to fabricate FeSe nanobridge. The scanning ion microscopy images of the fabricated FeSe nanobridges with (e) $W$ = 1250 nm, and (f) $W$ = 390 nm. The nanobridge in (f) was fabricated by reducing the width of the bridge in (e). (g) Temperature dependence of resistance for the two devices. The $I-V$	curves measured at different temperatures at zero field for the devices with (h) $W$ = 1250 nm, and (i) $W$ = 390 nm.}\label{}
\end{figure*}

Among the IBSs, FeSe has attracted significant interest due to its advantages for both probing the pairing mechanism \cite{Hanaguri2018SA} and its potential for practical applications \cite{SiWeidongNatComm}. Although its initial $T_{\rm{c}}$ is below 10 K \cite{HsuFongChiFeSediscovery}, it increases up to over 14 K with appropriate Te substitution \cite{SalesPRB}, 37 K under high pressure \cite{MedvedevNatMat,BurrardNatMat}, and 44 K by electro-chemical protonation \cite{MengyanPhysRevB.105.134506}. Furthermore, the monolayer of FeSe film grown on SrTiO$_3$ even shows a sign of superconductivity over 65 K \cite{GeNatMatter}. For applications, high quality Te-doped FeSe tapes with transport $J_{\rm{c}}$ over 10$^6$ A/cm$^2$ under self-field and over 10$^5$ A/cm$^2$ under 30 T at 4.2 K were already fabricated \cite{SiWeidongNatComm}. In this report, we provide experimental evidence that the depairing limit and the vortex-free state can be achieved through geometric constraint in a FeSe single crystal. The obtained transport $J_{\rm{c}}$ is enhanced by just reducing the width in an unique bridge, and its value can reach one order of magnitude larger than the depinning $J_{\rm{c}}$, indicating the obtained $J_{\rm{c}}$ is the depairing $J_{\rm{c}}$. Besides, the vortex-free state is found to be enhanced up to 1 kOe, due to the enhancement of $\mu_0H_{\rm{c1}}$.

\section{experiment}

High-quality FeSe single crystals were grown using the vapor transport method, as previously described in our reports \cite{SunPhysRevB.93.104502,Sun_sustreview,HouPNAS}. The nanobridge devices were fabricated using the focused ion beam (FIB) technique, employing the well-established "pickup" method commonly utilized in the preparation of samples for transmission electron microscopy (TEM) \cite{Mollreview,Miyazawa_2021}. Fig. 1(a) depicts the fabrication process of a lamella along the $c$ axis, achieved through FIB etching, resulting in a typical size of 10$\times$10 $\mu$m$^2$ and a thickness below 1 $\mu$m. Subsequently, the lamella was meticulously lifted using a quartz needle and transferred onto a silicon substrate with a contact-pattern. The lamella was then connected to the Au contact pads through Pt deposition. By utilizing FIB etching, a narrow bridge was created between two voltage terminals, in the $ab$ plane. The nanobridge's width can be further reduced by post-fabrication. This methodology allows us to obtain the transport $J_{\rm{c}}$ from nanobridges with varying widths at the same location of a single crystal, effectively eliminating any influence stemming from crystal quality. Schematic diagrams illustrating the fabrication process of the nanobridge device can be observed in Figs. 1(a)-1(d). Figs. 1(e) and 1(f) present scanning ion microscopy images of the nanobridges in the $ab$ plane, exhibiting widths of 1250 nm and 390 nm, respectively. The Pearl length of FeSe in the $ab$-plane is calculated as $\varLambda$ = 2$\lambda^2/h$ = 620 nm with $\lambda$= 450 nm, and $h$ = 650 nm \cite{SundepairingJc}. Hence, $W$ of the two devices correspond to $\sim$ 2$\varLambda$, and $\textless$ $\varLambda$.

\begin{figure}\center
	\includegraphics[width=8.5cm]{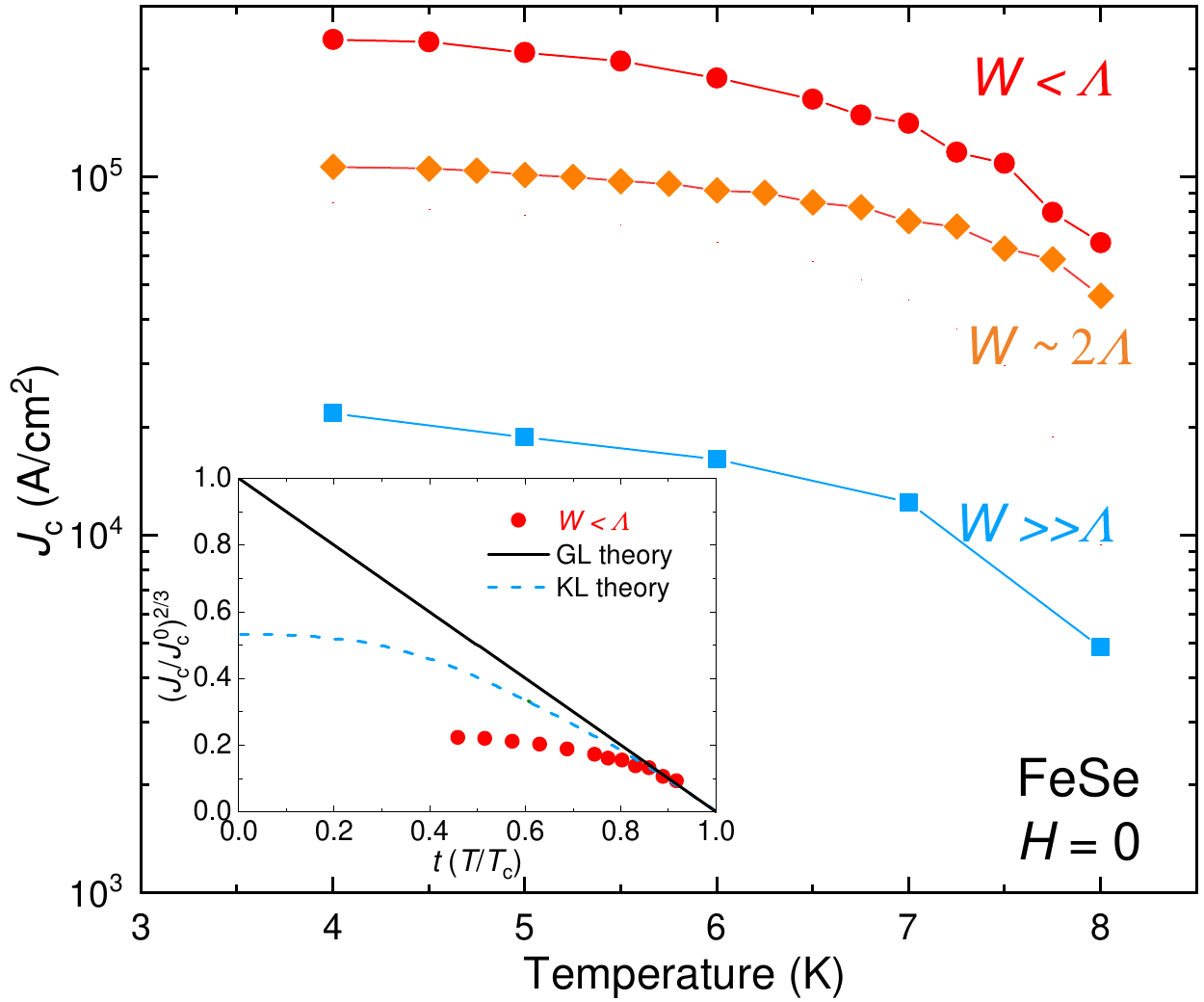}
	\caption{Temperature dependence of $J_{\rm{c}}$ under zero field for FeSe nanobridges with $W \gg \varLambda$, $W \sim 2\varLambda$, and $W \textless$ $\varLambda$. Inset shows the the reduced temperature ($t$ = $T$/$T_{\rm{c}}$) dependence of current density density, normalized to the extrapolated value $J_{\rm{c}}$(0) for the device with $W \textless$ $\varLambda$. The solid and the dashed lines represent the results from the GL theory and KL theory, respectively}\label{}
\end{figure}

Resistance measurements were performed using a standard four-probe technique. The device with a bridge width of $W$ = 1250 nm exhibits a $T_{\rm{c}}$ $\sim$ 8 K, comparable to that of bulk crystals, with a sharp transition width of less than 0.5 K (see Fig. 1(g)). Upon reducing the bridge width to 390 nm, the superconducting transition becomes slightly broader, with $T_{\rm{c}}$ (zero) decreasing to around 7.5 K. The suppression of $T_{\rm{c}}$ is attributed to damage induced by Ga$^+$ ion implantation during fabrication, a commonly observed effect in FIB processing \cite{MizukoshiPhysRevB.110.104501}. To obtain $I$-$V$ characteristics below $T_{\rm{c}}$, measurements were performed up to 5 T using pulsed currents applied to the nanobridges. The Keithley Delta Pulse System was utilized to generate the pulse current. Rectangular current pulses with a duration of 100 $\mu$s were applied to the sample at intervals of 3 s (duty ratio $\sim$ 3.3 $\times$ 10$^{-5}$) to minimize any heating effects. The voltage drop across the bridge was then integrated over a period of 55 $\mu$s. For the $I$-$V$ measurements under magnetic field, the bridge was zero-field cooled to the target temperature before applying fields. The $I$-$V$ curves for the two devices measured at various temperatures are presented in Figs. 1(h) and 1(i), respectively. A criterion of 100 $\mu$V was used to define the critical current ($I_{\rm{c}}$). Here, we would like to highlight that the $I-V$ transition near $T_{\rm{c}}$ in the device with $W$ = 390 nm appears slightly broader due to the extended superconducting transition, as discussed earlier. However, at lower temperatures, the transition becomes sharper, minimizing the impact of the chosen criterion on the estimated $I_{\rm{c}}$. For instance, at 4 K, using a 10 $\mu$V criterion yields an $I_{\rm{c}}$ value of approximately 0.55 mA, only about 5\% lower than that obtained with a 100 $\mu$V criterion. Since our study primarily focuses on low-temperature behavior, this slight broadening of the $I-V$ transition does not affect our main conclusions. According to the 100 $\mu$V criterion, the error bars were estimated to be less than 5\% based on the applied current step of 0.01 mA, and are not displayed because their magnitude is smaller than the symbol size.

\section{results and discussion}

\begin{figure}\center
	\includegraphics[width=8.5cm]{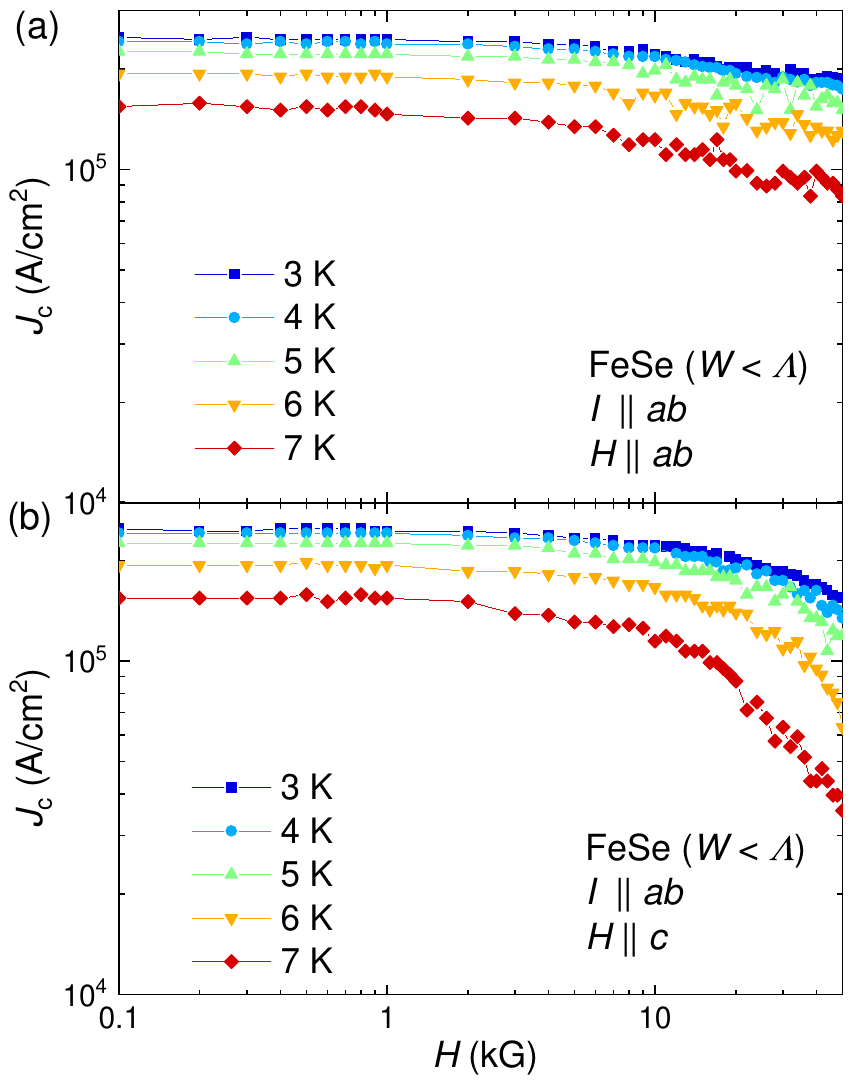}
	\caption{Magnetic field dependence of $J_{\rm{c}}$ at different temperatures obtained from the $I-V$ measurements in the FeSe nanobridge with $W \textless$ $\varLambda$ for (a) $H \parallel ab$, and (b) $H \parallel c$.}\label{}
\end{figure}

Figure 2 illustrates the temperature dependence of $J_{\rm{c}}$ under zero field for bridges of varying width. In the following of the article, we use the magnitude relationship between the width ($W$) and Pearl length ($\varLambda$ = 2$\lambda^2/h$, where $\lambda$ is the penetration depth) to referring the bulk and the two fabricated devices as $W\gg\varLambda$, $W$ $\sim$ 2$\varLambda$, and $W \textless \varLambda$, respectively. When the width is significantly larger than $\varLambda$ (bulk crystal), determining $J_{\rm{c}}$ from the $I-V$ curve measurements becomes challenging due to the requirement of a substantial applied current to reach $I_{\rm{c}}$, which often leads to substantial heating effects. Consequently, $J_{\rm{c}}$ for $W \gg \varLambda$ is estimated using the magnetization hysteresis loops (MHLs) based on the Bean model \cite{supplement}, which captures the vortex depinning \cite{SunPhysRevB.92.144509}. It has been extensively established that the $J_{\rm{c}}$ calculated from the MHLs in a bulk single crystal exhibits close agreement with the values obtained from the $I-V$ curve measurements \cite{ProzorovPRBBaCoJc}. For the two nanobridges with $W \sim 2\varLambda$ and $W \textless$ $\varLambda$, $J_{\rm{c}}$ is directly obtained from the $I-V$ measurements.  Notably, the $J_{\rm{c}}$ values for $I \parallel ab$ significantly increase as the bridge width decreases. For $W \gg \varLambda$, $J_{\rm{c}}$ is approximately 2 $\times$ 10$^4$ A/cm$^2$ at 4 K. It rises to around 5 $\times$ 10$^4$ A/cm$^2$ for $W \sim 2\varLambda$, and further surpasses 2 $\times$ 10$^5$ A/cm$^2$ for $W \textless$ $\varLambda$. The increase of $J_{\rm{c}}$ by reducing the bridge's width has been also observed in YBCO, and can be explained by the enhancement of the homogeneity of $J_{\rm{c}}$ \cite{NawazPhysRevLettYBCOJdp}. $J_{\rm{c}}$ for $W \textless \varLambda$ is more than one order of magnitude greater than the depinning $J_{\rm{c}}$, indicating the achievement of the depairing limit. We would like to emphasize that the notable enhancement in $J_{\rm{c}}$ is exclusively achieved by reducing the width of the nanobridge within a specific location of an unique crystal. This enhancement cannot be accounted for solely by the depinning mechanism and can only be attributed to the transition from depinning to depairing.

Based on the Ginzburg-Landau (GL) theory, $J_{\rm{c}}$ close to $T_{\rm{c}}$ can be described as a function of the reduced temperature $t$ = $T$/$T_{\rm{c}}$ by the formula $J_{\rm{c}}$($t$)=$J_{\rm{c}}$(0)(1-$t$)$^{3/2}$ \cite{Tinkham}. To compare our experimental data with the theory, $J_{\rm{c}}^{\rm{2/3}}$ close to $T_{\rm{c}}$ was linearly fitted, and extrapolated to $t$ = 0 to extract $J_{\rm{c}}$(0). The reduced critical current density $J_{\rm{c}}$($t$)/$J_{\rm{c}}$(0) are shown in the inset of Fig. 2, together with the theoretical results from the GL-theory (solid line) and Kupriyanov-Lukichev (KL)-theory (dashed line) \cite{KLtheory}. In KL theory, the depairing $J_{\rm{c}}$ was numerically calculated from the Eilenberger equations by assuming that the velocity of supercurrent is proportional to a phase gradient of the SC order parameter. Clearly, the measured depairing $J_{\rm{c}}$ is much smaller than the theoretical ones. The over-estimated theoretical depairing $J_{\rm{c}}$ may be due to the special characteristics of FeSe, which hosts tiny Fermi surface with multi-bands and resides in the crossover regime of BCS to BEC \cite{Kasahara18112014,ShibauchiFeSereview}, leading to a stronger electronic correlations than in a simple BCS superconductor considered in the theories.     

\begin{figure}\center
	\includegraphics[width=8.5cm]{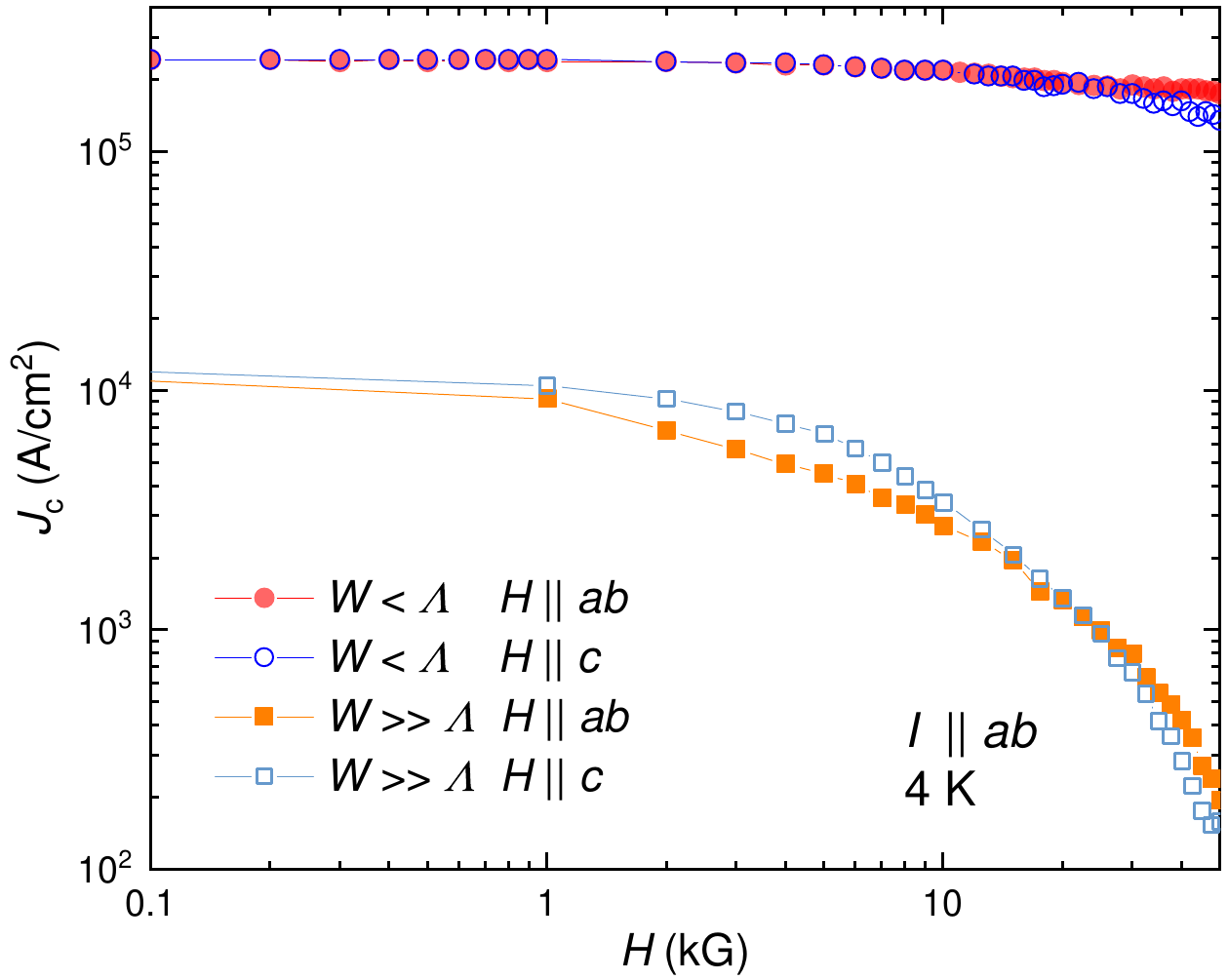}
	\caption{Comparison of the field dependence of $J_{\rm{c}}$ at 4 K for FeSe with $W \textless$ $\varLambda$ and $W \gg \varLambda$ for both $H \parallel ab$, and $H \parallel c$.}\label{}
\end{figure}

To future investigate the depairing $J_{\rm{c}}$, we conducted $I-V$ measurements under varying magnetic fields on the nanobridge within the $ab$ plane, where the width $W \textless \varLambda$. These measurements were performed for magnetic fields oriented parallel to both the $ab$ plane ($H \parallel ab$) and the $c$ axis ($H \parallel c$). The field dependence of $J_{\rm{c}}$ measured over a temperature range spanning from 3 K to 7 K, for both $H \parallel ab$ and $H \parallel c$ are illustrated in Figs. 3(a) and 3(b), respectively. Intriguingly, the value of $J_{\rm{c}}$ remains almost unchanged in the presence of magnetic fields up to 1 kOe, experiencing only a slight reduction until approximately 10 kOe. However, beyond this point, $J_{\rm{c}}$ begins to exhibit a more pronounced decline with increasing magnetic field strength. This robust behavior of $J_{\rm{c}}$ against magnetic fields, up to 1 kOe, stands in stark contrast to the depinning scenario, where $J_{\rm{c}}$ is significantly more responsive to changes in magnetic field. The comparison of field dependence of $J_{\rm{c}}$ with $W \textless$ $\varLambda$ (depairing) and $W \gg \varLambda$ (depinning) is directly shown in Fig. 4. 

In the case of depinning, $J_{\rm{c}}$ decreases quickly with the magnetic field due to the entry of vortices when the field exceeds $\mu_0H_{\rm{c1}}$. After the entry of vortices, the decrease in $J_{\rm{c}}$ depends on the movement of vortices, i.e., the pinning of the crystals. For most bulk cuprates and IBSs, the magnitude of $\mu_0H_{\rm{c1}}$ is typically in the range of 20 to 200 Oe \cite{Nideroest1998,JoshiPhysRevApplied.11.014035,Song_2011LiFeSepenetrationdepth}. In FeSe, it was reported to be approximately 30 Oe at 0 K \cite{AbdelFeSeHc1PhysRevB.88.174512}. On the other hand, the robust $J_{\rm{c}}$ against fields up to around 1 kOe (see Fig. 3) indicates that the vortex-free state can be sustained up to approximately 1 kOe in the FeSe nanobridge with $W \textless \varLambda$. In thin layered superconductors,  $\mu_0H_{\rm{c1}}$ is dependent on its width when $W$ is comparable to $\lambda$. Based on Abrikosov's calculation \cite{AbrikosovHc1}, $\mu_0H_{\rm{c1}}$ can be expressed as
$\begin{aligned}
	\mu_0H_{c1}& =\frac{\Phi_0}{4\pi\lambda^2}\Bigg(\log\kappa+0.081-2\sum_{n=1}^\infty(-1)^{n+1}K_0\Bigg(\frac{Wn}\lambda\Bigg)\Bigg) \\
	&\left(1 - \mathrm{sech}\!\left(\frac W{2\lambda}\right)\right)^{-1},
\end{aligned}$          
where $\Phi_0$ is the flux quantum, $K_0$ is the zero-order modified Bessel function of the second kind, and $\kappa$=$\lambda/\xi$. In the case of FeSe, $\mu_0H_{\rm{c1}}$ is calculated and plotted in Fig. 5 using $\lambda$ $\sim$ 450 nm, and $\xi$ $\sim$ 4.3 nm \cite{SundepairingJc}.

\begin{figure}\center
	\includegraphics[width=8.5cm]{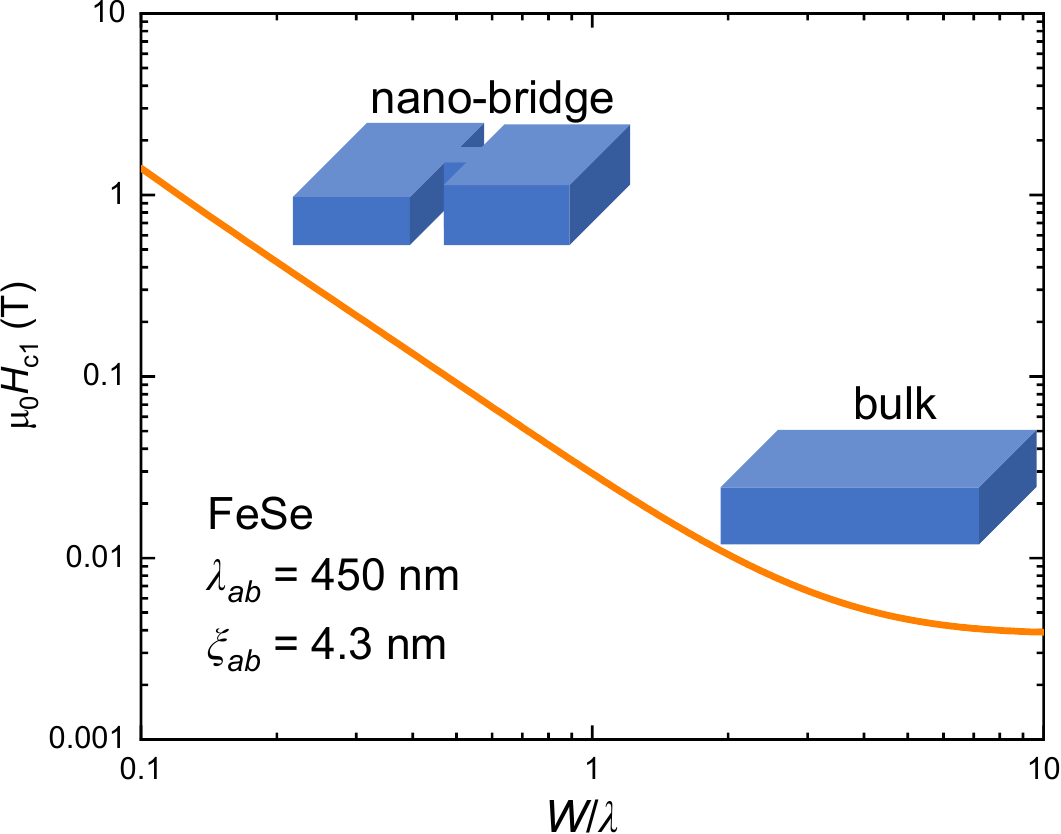}
	\caption{Bridge width dependence of $\mu_0H_{\rm{c1}}$ in FeSe calculated in the range of 0.1 $\leq$ $W/\lambda$ $\leq$ 10 based on Abrikosov's formula.}\label{}
\end{figure}

As shown in Fig. 5, $\mu_0H_{\rm{c1}}$ approaches 30 Oe when $W$ is much larger than $\lambda$, which is consistent with experimental reports. As $W$ decreases, $\mu_0H_{\rm{c1}}$ gradually increases. When $W$/$\lambda$ is approximately 2, $\mu_0H_{\rm{c1}}$ exceeds 100 Oe. With future reduction of $W$, $\mu_0H_{\rm{c1}}$ increases almost linearly with $W$/$\lambda$ in the double-log plot. It can exceed 1 kOe when $W$/$\lambda$ = 0.5, and even surpass 10 kOe when $W$/$\lambda$=0.1. Clearly, the theoretical calculation is consistent with our experimental observation that the vortex-free state (i.e. $H$ $\textless$ $H_{\rm{c1}}$) can be extended to higher fields by decreasing the $W$ of the sample. On the other hand, the calculated $\mu_0H_{\rm{c1}}$ is still smaller than the values observed in Fig. 3. The calculation is based on the assumption that vortex trapping in a strip is absolutely stable, with the relevant terms in the free energy equated to zero \cite{AbrikosovHc1}. Under this assumption, the increase of $\mu_0H_{\rm{c1}}$ saturates in the limit of small $W$. As a result, the calculated $\mu_0H_{\rm{c1}}$ underestimates the actual value in this regime. Such behavior has indeed been observed in earlier studies of YBa$_2$Cu$_3$O$_{7-d}$ nanowires \cite{Victornanolett}. To address this discrepancy, Kuit $et al$. \cite{KuitPhysRevB.77.134504} proposed an alternative model that incorporates the generation of vortex–antivortex pairs in the strip. In the current case, we considered that the demagnetization effect arising from the sample geometry \cite{Galluzzicondmat8040091} and the extrinsic pinning at the disordered surface ( thickness $\sim$ 5 nm) introduced by FIB milling \cite{Mollreview} could hinder the entry of magnetic flux and thus lead to an increase in $\mu_0H_{\rm{c1}}$. Since it is extremely challenging to incorporate both effects into the calculation of $\mu_0H_{\rm{c1}}$, experimental approaches—such as varying the aspect ratio of the microbridges and adjusting the FIB current during fabrication—may provide valuable insights.

\section{conclusions}

We systematically studied the geometric constraint effect on the $J_{\rm{c}}$ of FeSe superconductors. $J_{\rm{c}}$ is found to be continuously enhanced by only reducing the width of an unique bridge, finally reaching a magnitude more than one order larger than the depinning $J_{\rm{c}}$, which evidence the achieving of the depairing limit. $J_{\rm{c}}$ is also found to be robust against magnetic field up to 1 kOe when its width is smaller than $\lambda$. Based on the Abrikosov's calculation, $\mu_0H_{\rm{c1}}$ is enhanced by reducing the width of the bridge when it is comparable to $\lambda$. Therefore, the field-robust $J_{\rm{c}}$ is the evidence of the vortex-free state. Our study provides a promising way to study the supercurrent related physics and the application free from vortex-motion linked noise. Future observations using scanning tunneling microscope (STM) or magnetic force microscope (MFM) could directly reveal the expulsion of vortices induced by geometric confinement.

\acknowledgements
The present work was partly supported by the National Key R\&D Program of China (Grant No. 2024YFA1408400), the National Natural Science Foundation of China (Grants No. 12374136), the CAS Superconducting Research Project under Grant No. [SCZX-0101], and KAKENHI (20H05164) from JSPS. FIB microfabrication performed in this work was supported by Center for Instrumental Analysis, College of Science and Engineering, Aoyama Gakuin University.

S.Y. and Y.X. contributed equally to this paper.


\bibliography{references}

\begin{thebibliography}{38}%
\makeatletter
\providecommand \@ifxundefined [1]{%
 \@ifx{#1\undefined}
}%
\providecommand \@ifnum [1]{%
 \ifnum #1\expandafter \@firstoftwo
 \else \expandafter \@secondoftwo
 \fi
}%
\providecommand \@ifx [1]{%
 \ifx #1\expandafter \@firstoftwo
 \else \expandafter \@secondoftwo
 \fi
}%
\providecommand \natexlab [1]{#1}%
\providecommand \enquote  [1]{``#1''}%
\providecommand \bibnamefont  [1]{#1}%
\providecommand \bibfnamefont [1]{#1}%
\providecommand \citenamefont [1]{#1}%
\providecommand \href@noop [0]{\@secondoftwo}%
\providecommand \href [0]{\begingroup \@sanitize@url \@href}%
\providecommand \@href[1]{\@@startlink{#1}\@@href}%
\providecommand \@@href[1]{\endgroup#1\@@endlink}%
\providecommand \@sanitize@url [0]{\catcode `\\12\catcode `\$12\catcode
  `\&12\catcode `\#12\catcode `\^12\catcode `\_12\catcode `\%12\relax}%
\providecommand \@@startlink[1]{}%
\providecommand \@@endlink[0]{}%
\providecommand \url  [0]{\begingroup\@sanitize@url \@url }%
\providecommand \@url [1]{\endgroup\@href {#1}{\urlprefix }}%
\providecommand \urlprefix  [0]{URL }%
\providecommand \Eprint [0]{\href }%
\providecommand \doibase [0]{https://doi.org/}%
\providecommand \selectlanguage [0]{\@gobble}%
\providecommand \bibinfo  [0]{\@secondoftwo}%
\providecommand \bibfield  [0]{\@secondoftwo}%
\providecommand \translation [1]{[#1]}%
\providecommand \BibitemOpen [0]{}%
\providecommand \bibitemStop [0]{}%
\providecommand \bibitemNoStop [0]{.\EOS\space}%
\providecommand \EOS [0]{\spacefactor3000\relax}%
\providecommand \BibitemShut  [1]{\csname bibitem#1\endcsname}%
\let\auto@bib@innerbib\@empty
\bibitem [{\citenamefont {Tinkham}(1996)}]{Tinkham}%
  \BibitemOpen
  \bibfield  {author} {\bibinfo {author} {\bibfnamefont {M.}~\bibnamefont
  {Tinkham}},\ }\href@noop {} {\emph {\bibinfo {title} {Introduction to
  superconductivity}}}\ (\bibinfo  {publisher} {Courier Corporation},\ \bibinfo
  {address} {New York},\ \bibinfo {year} {1996})\BibitemShut {NoStop}%
\bibitem [{\citenamefont {Dew-Hughes}(2001)}]{Dew-HughesJcReview}%
  \BibitemOpen
  \bibfield  {author} {\bibinfo {author} {\bibfnamefont {D.}~\bibnamefont
  {Dew-Hughes}},\ }\bibfield  {title} {\bibinfo {title} {The critical current
  of superconductors: an historical review},\ }\href
  {https://doi.org/10.1063/1.1401180} {\bibfield  {journal} {\bibinfo
  {journal} {Low Temp. Phys.}\ }\textbf {\bibinfo {volume} {27}},\ \bibinfo
  {pages} {713} (\bibinfo {year} {2001})}\BibitemShut {NoStop}%
\bibitem [{\citenamefont {Skocpol}(1976)}]{SkocpolPhysRevB.14.1045}%
  \BibitemOpen
  \bibfield  {author} {\bibinfo {author} {\bibfnamefont {W.~J.}\ \bibnamefont
  {Skocpol}},\ }\href {https://doi.org/10.1103/PhysRevB.14.1045} {\bibfield
  {journal} {\bibinfo  {journal} {Phys. Rev. B}\ }\textbf {\bibinfo {volume}
  {14}},\ \bibinfo {pages} {1045} (\bibinfo {year} {1976})}\BibitemShut
  {NoStop}%
\bibitem [{\citenamefont {Blatter}\ \emph {et~al.}(1994)\citenamefont
  {Blatter}, \citenamefont {Feigel'man}, \citenamefont {Geshkenbein},
  \citenamefont {Larkin},\ and\ \citenamefont {Vinokur}}]{Blatterreview}%
  \BibitemOpen
  \bibfield  {author} {\bibinfo {author} {\bibfnamefont {G.}~\bibnamefont
  {Blatter}}, \bibinfo {author} {\bibfnamefont {M.~V.}\ \bibnamefont
  {Feigel'man}}, \bibinfo {author} {\bibfnamefont {V.~B.}\ \bibnamefont
  {Geshkenbein}}, \bibinfo {author} {\bibfnamefont {A.~I.}\ \bibnamefont
  {Larkin}},\ and\ \bibinfo {author} {\bibfnamefont {V.~M.}\ \bibnamefont
  {Vinokur}},\ }\href {https://link.aps.org/doi/10.1103/RevModPhys.66.1125}
  {\bibfield  {journal} {\bibinfo  {journal} {Rev. Mod. Phys.}\ }\textbf
  {\bibinfo {volume} {66}},\ \bibinfo {pages} {1125} (\bibinfo {year}
  {1994})}\BibitemShut {NoStop}%
\bibitem [{\citenamefont {Tsuei}\ and\ \citenamefont
  {Kirtley}(2000)}]{TsueiRevModPhys.72.969}%
  \BibitemOpen
  \bibfield  {author} {\bibinfo {author} {\bibfnamefont {C.~C.}\ \bibnamefont
  {Tsuei}}\ and\ \bibinfo {author} {\bibfnamefont {J.~R.}\ \bibnamefont
  {Kirtley}},\ }\href {https://doi.org/10.1103/RevModPhys.72.969} {\bibfield
  {journal} {\bibinfo  {journal} {Rev. Mod. Phys.}\ }\textbf {\bibinfo {volume}
  {72}},\ \bibinfo {pages} {969} (\bibinfo {year} {2000})}\BibitemShut
  {NoStop}%
\bibitem [{\citenamefont {Nawaz}\ \emph {et~al.}(2013)\citenamefont {Nawaz},
  \citenamefont {Arpaia}, \citenamefont {Lombardi},\ and\ \citenamefont
  {Bauch}}]{NawazPhysRevLettYBCOJdp}%
  \BibitemOpen
  \bibfield  {author} {\bibinfo {author} {\bibfnamefont {S.}~\bibnamefont
  {Nawaz}}, \bibinfo {author} {\bibfnamefont {R.}~\bibnamefont {Arpaia}},
  \bibinfo {author} {\bibfnamefont {F.}~\bibnamefont {Lombardi}},\ and\
  \bibinfo {author} {\bibfnamefont {T.}~\bibnamefont {Bauch}},\ }\href
  {https://link.aps.org/doi/10.1103/PhysRevLett.110.167004} {\bibfield
  {journal} {\bibinfo  {journal} {Phys. Rev. Lett.}\ }\textbf {\bibinfo
  {volume} {110}},\ \bibinfo {pages} {167004} (\bibinfo {year}
  {2013})}\BibitemShut {NoStop}%
\bibitem [{\citenamefont {Clem}\ and\ \citenamefont
  {Berggren}(2011)}]{ClemPearllength}%
  \BibitemOpen
  \bibfield  {author} {\bibinfo {author} {\bibfnamefont {J.~R.}\ \bibnamefont
  {Clem}}\ and\ \bibinfo {author} {\bibfnamefont {K.~K.}\ \bibnamefont
  {Berggren}},\ }\href {https://doi.org/10.1103/PhysRevB.84.174510} {\bibfield
  {journal} {\bibinfo  {journal} {Phys. Rev. B}\ }\textbf {\bibinfo {volume}
  {84}},\ \bibinfo {pages} {174510} (\bibinfo {year} {2011})}\BibitemShut
  {NoStop}%
\bibitem [{\citenamefont {Li}\ \emph {et~al.}(2013)\citenamefont {Li},
  \citenamefont {Yuan}, \citenamefont {Yuan}, \citenamefont {Ge}, \citenamefont
  {Li}, \citenamefont {Feng}, \citenamefont {Pereira}, \citenamefont {Ishii},
  \citenamefont {Hatano}, \citenamefont {Silhanek}, \citenamefont {Chibotaru},
  \citenamefont {Vanacken}, \citenamefont {Yamaura}, \citenamefont {Wang},
  \citenamefont {Takayama-Muromachi},\ and\ \citenamefont
  {Moshchalkov}}]{LiJunAPLdeparingJcBaK122}%
  \BibitemOpen
  \bibfield  {author} {\bibinfo {author} {\bibfnamefont {J.}~\bibnamefont
  {Li}}, \bibinfo {author} {\bibfnamefont {J.}~\bibnamefont {Yuan}}, \bibinfo
  {author} {\bibfnamefont {Y.-H.}\ \bibnamefont {Yuan}}, \bibinfo {author}
  {\bibfnamefont {J.-Y.}\ \bibnamefont {Ge}}, \bibinfo {author} {\bibfnamefont
  {M.-Y.}\ \bibnamefont {Li}}, \bibinfo {author} {\bibfnamefont {H.-L.}\
  \bibnamefont {Feng}}, \bibinfo {author} {\bibfnamefont {P.~J.}\ \bibnamefont
  {Pereira}}, \bibinfo {author} {\bibfnamefont {A.}~\bibnamefont {Ishii}},
  \bibinfo {author} {\bibfnamefont {T.}~\bibnamefont {Hatano}}, \bibinfo
  {author} {\bibfnamefont {A.~V.}\ \bibnamefont {Silhanek}}, \bibinfo {author}
  {\bibfnamefont {L.~F.}\ \bibnamefont {Chibotaru}}, \bibinfo {author}
  {\bibfnamefont {J.}~\bibnamefont {Vanacken}}, \bibinfo {author}
  {\bibfnamefont {K.}~\bibnamefont {Yamaura}}, \bibinfo {author} {\bibfnamefont
  {H.-B.}\ \bibnamefont {Wang}}, \bibinfo {author} {\bibfnamefont
  {E.}~\bibnamefont {Takayama-Muromachi}},\ and\ \bibinfo {author}
  {\bibfnamefont {V.~V.}\ \bibnamefont {Moshchalkov}},\ }\href
  {https://doi.org/10.1063/1.4818127} {\bibfield  {journal} {\bibinfo
  {journal} {Appl. Phys. Lett.}\ }\textbf {\bibinfo {volume} {103}},\ \bibinfo
  {pages} {062603} (\bibinfo {year} {2013})}\BibitemShut {NoStop}%
\bibitem [{\citenamefont {Sun}\ \emph {et~al.}(2020)\citenamefont {Sun},
  \citenamefont {Ohnuma}, \citenamefont {Ayukawa}, \citenamefont {Noji},
  \citenamefont {Koike}, \citenamefont {Tamegai},\ and\ \citenamefont
  {Kitano}}]{SundepairingJc}%
  \BibitemOpen
  \bibfield  {author} {\bibinfo {author} {\bibfnamefont {Y.}~\bibnamefont
  {Sun}}, \bibinfo {author} {\bibfnamefont {H.}~\bibnamefont {Ohnuma}},
  \bibinfo {author} {\bibfnamefont {S.-y.}\ \bibnamefont {Ayukawa}}, \bibinfo
  {author} {\bibfnamefont {T.}~\bibnamefont {Noji}}, \bibinfo {author}
  {\bibfnamefont {Y.}~\bibnamefont {Koike}}, \bibinfo {author} {\bibfnamefont
  {T.}~\bibnamefont {Tamegai}},\ and\ \bibinfo {author} {\bibfnamefont
  {H.}~\bibnamefont {Kitano}},\ }\href
  {https://doi.org/10.1103/PhysRevB.101.134516} {\bibfield  {journal} {\bibinfo
   {journal} {Phys. Rev. B}\ }\textbf {\bibinfo {volume} {101}},\ \bibinfo
  {pages} {134516} (\bibinfo {year} {2020})}\BibitemShut {NoStop}%
\bibitem [{\citenamefont {Mizukoshi}\ \emph {et~al.}(2024)\citenamefont
  {Mizukoshi}, \citenamefont {Jimbo}, \citenamefont {Park}, \citenamefont
  {Sun}, \citenamefont {Tamegai},\ and\ \citenamefont
  {Kitano}}]{MizukoshiPhysRevB.110.104501}%
  \BibitemOpen
  \bibfield  {author} {\bibinfo {author} {\bibfnamefont {Y.}~\bibnamefont
  {Mizukoshi}}, \bibinfo {author} {\bibfnamefont {K.}~\bibnamefont {Jimbo}},
  \bibinfo {author} {\bibfnamefont {A.}~\bibnamefont {Park}}, \bibinfo {author}
  {\bibfnamefont {Y.}~\bibnamefont {Sun}}, \bibinfo {author} {\bibfnamefont
  {T.}~\bibnamefont {Tamegai}},\ and\ \bibinfo {author} {\bibfnamefont
  {H.}~\bibnamefont {Kitano}},\ }\href
  {https://doi.org/10.1103/PhysRevB.110.104501} {\bibfield  {journal} {\bibinfo
   {journal} {Phys. Rev. B}\ }\textbf {\bibinfo {volume} {110}},\ \bibinfo
  {pages} {104501} (\bibinfo {year} {2024})}\BibitemShut {NoStop}%
\bibitem [{\citenamefont {Rouco}\ \emph {et~al.}(2019)\citenamefont {Rouco},
  \citenamefont {Navau}, \citenamefont {Del-Valle}, \citenamefont {Massarotti},
  \citenamefont {Papari}, \citenamefont {Stornaiuolo}, \citenamefont
  {Obradors}, \citenamefont {Puig}, \citenamefont {Tafuri}, \citenamefont
  {Sanchez},\ and\ \citenamefont {Palau}}]{Victornanolett}%
  \BibitemOpen
  \bibfield  {author} {\bibinfo {author} {\bibfnamefont {V.}~\bibnamefont
  {Rouco}}, \bibinfo {author} {\bibfnamefont {C.}~\bibnamefont {Navau}},
  \bibinfo {author} {\bibfnamefont {N.}~\bibnamefont {Del-Valle}}, \bibinfo
  {author} {\bibfnamefont {D.}~\bibnamefont {Massarotti}}, \bibinfo {author}
  {\bibfnamefont {G.~P.}\ \bibnamefont {Papari}}, \bibinfo {author}
  {\bibfnamefont {D.}~\bibnamefont {Stornaiuolo}}, \bibinfo {author}
  {\bibfnamefont {X.}~\bibnamefont {Obradors}}, \bibinfo {author}
  {\bibfnamefont {T.}~\bibnamefont {Puig}}, \bibinfo {author} {\bibfnamefont
  {F.}~\bibnamefont {Tafuri}}, \bibinfo {author} {\bibfnamefont
  {A.}~\bibnamefont {Sanchez}},\ and\ \bibinfo {author} {\bibfnamefont
  {A.}~\bibnamefont {Palau}},\ }\href
  {https://doi.org/10.1021/acs.nanolett.9b01693} {\bibfield  {journal}
  {\bibinfo  {journal} {Nano Letters}\ }\textbf {\bibinfo {volume} {19}},\
  \bibinfo {pages} {4174} (\bibinfo {year} {2019})}\BibitemShut {NoStop}%
\bibitem [{\citenamefont {Kuit}\ \emph {et~al.}(2008)\citenamefont {Kuit},
  \citenamefont {Kirtley}, \citenamefont {van~der Veur}, \citenamefont
  {Molenaar}, \citenamefont {Roesthuis}, \citenamefont {Troeman}, \citenamefont
  {Clem}, \citenamefont {Hilgenkamp}, \citenamefont {Rogalla},\ and\
  \citenamefont {Flokstra}}]{KuitPhysRevB.77.134504}%
  \BibitemOpen
  \bibfield  {author} {\bibinfo {author} {\bibfnamefont {K.~H.}\ \bibnamefont
  {Kuit}}, \bibinfo {author} {\bibfnamefont {J.~R.}\ \bibnamefont {Kirtley}},
  \bibinfo {author} {\bibfnamefont {W.}~\bibnamefont {van~der Veur}}, \bibinfo
  {author} {\bibfnamefont {C.~G.}\ \bibnamefont {Molenaar}}, \bibinfo {author}
  {\bibfnamefont {F.~J.~G.}\ \bibnamefont {Roesthuis}}, \bibinfo {author}
  {\bibfnamefont {A.~G.~P.}\ \bibnamefont {Troeman}}, \bibinfo {author}
  {\bibfnamefont {J.~R.}\ \bibnamefont {Clem}}, \bibinfo {author}
  {\bibfnamefont {H.}~\bibnamefont {Hilgenkamp}}, \bibinfo {author}
  {\bibfnamefont {H.}~\bibnamefont {Rogalla}},\ and\ \bibinfo {author}
  {\bibfnamefont {J.}~\bibnamefont {Flokstra}},\ }\bibfield  {title} {\bibinfo
  {title} {Vortex trapping and expulsion in thin-film
  ${\text{yba}}_{2}{\text{cu}}_{3}{\text{o}}_{7\ensuremath{-}\ensuremath{\delta}}$
  strips},\ }\href {https://doi.org/10.1103/PhysRevB.77.134504} {\bibfield
  {journal} {\bibinfo  {journal} {Phys. Rev. B}\ }\textbf {\bibinfo {volume}
  {77}},\ \bibinfo {pages} {134504} (\bibinfo {year} {2008})}\BibitemShut
  {NoStop}%
\bibitem [{\citenamefont {Hosono}\ and\ \citenamefont
  {Kuroki}(2015)}]{Hosono2015}%
  \BibitemOpen
  \bibfield  {author} {\bibinfo {author} {\bibfnamefont {H.}~\bibnamefont
  {Hosono}}\ and\ \bibinfo {author} {\bibfnamefont {K.}~\bibnamefont
  {Kuroki}},\ }\href {https://doi.org/10.1016/j.physc.2015.02.020} {\bibfield
  {journal} {\bibinfo  {journal} {Physica C: Superconductivity and its
  Applications}\ }\textbf {\bibinfo {volume} {514}},\ \bibinfo {pages} {399}
  (\bibinfo {year} {2015})}\BibitemShut {NoStop}%
\bibitem [{\citenamefont {Hanaguri}\ \emph {et~al.}(2018)\citenamefont
  {Hanaguri}, \citenamefont {Kasahara}, \citenamefont {B{\"o}hmer},
  \citenamefont {Hardy}, \citenamefont {Meingast}, \citenamefont {Wolf},
  \citenamefont {Shibauchi},\ and\ \citenamefont {Matsuda}}]{Hanaguri2018SA}%
  \BibitemOpen
  \bibfield  {author} {\bibinfo {author} {\bibfnamefont {T.}~\bibnamefont
  {Hanaguri}}, \bibinfo {author} {\bibfnamefont {S.}~\bibnamefont {Kasahara}},
  \bibinfo {author} {\bibfnamefont {A.~E.}\ \bibnamefont {B{\"o}hmer}},
  \bibinfo {author} {\bibfnamefont {F.}~\bibnamefont {Hardy}}, \bibinfo
  {author} {\bibfnamefont {C.}~\bibnamefont {Meingast}}, \bibinfo {author}
  {\bibfnamefont {T.}~\bibnamefont {Wolf}}, \bibinfo {author} {\bibfnamefont
  {T.}~\bibnamefont {Shibauchi}},\ and\ \bibinfo {author} {\bibfnamefont
  {Y.}~\bibnamefont {Matsuda}},\ }\href
  {https://doi.org/10.1126/sciadv.aar6419} {\bibfield  {journal} {\bibinfo
  {journal} {Science Advances}\ }\textbf {\bibinfo {volume} {4}},\ \bibinfo
  {pages} {eaar6419} (\bibinfo {year} {2018})}\BibitemShut {NoStop}%
\bibitem [{\citenamefont {Si}\ \emph {et~al.}(2013)\citenamefont {Si},
  \citenamefont {Han}, \citenamefont {Shi}, \citenamefont {Ehrlich},
  \citenamefont {Jaroszynski}, \citenamefont {Goyal},\ and\ \citenamefont
  {Li}}]{SiWeidongNatComm}%
  \BibitemOpen
  \bibfield  {author} {\bibinfo {author} {\bibfnamefont {W.}~\bibnamefont
  {Si}}, \bibinfo {author} {\bibfnamefont {S.~J.}\ \bibnamefont {Han}},
  \bibinfo {author} {\bibfnamefont {X.}~\bibnamefont {Shi}}, \bibinfo {author}
  {\bibfnamefont {S.~N.}\ \bibnamefont {Ehrlich}}, \bibinfo {author}
  {\bibfnamefont {J.}~\bibnamefont {Jaroszynski}}, \bibinfo {author}
  {\bibfnamefont {A.}~\bibnamefont {Goyal}},\ and\ \bibinfo {author}
  {\bibfnamefont {Q.}~\bibnamefont {Li}},\ }\href
  {https://www.nature.com/articles/ncomms2337} {\bibfield  {journal} {\bibinfo
  {journal} {Nat. Commun.}\ }\textbf {\bibinfo {volume} {4}},\ \bibinfo {pages}
  {1347} (\bibinfo {year} {2013})}\BibitemShut {NoStop}%
\bibitem [{\citenamefont {Hsu}\ \emph {et~al.}(2008)\citenamefont {Hsu},
  \citenamefont {Luo}, \citenamefont {Yeh}, \citenamefont {Chen}, \citenamefont
  {Huang}, \citenamefont {Wu}, \citenamefont {Lee}, \citenamefont {Huang},
  \citenamefont {Chu}, \citenamefont {Yan},\ and\ \citenamefont
  {Wu}}]{HsuFongChiFeSediscovery}%
  \BibitemOpen
  \bibfield  {author} {\bibinfo {author} {\bibfnamefont {F.~C.}\ \bibnamefont
  {Hsu}}, \bibinfo {author} {\bibfnamefont {J.~Y.}\ \bibnamefont {Luo}},
  \bibinfo {author} {\bibfnamefont {K.~W.}\ \bibnamefont {Yeh}}, \bibinfo
  {author} {\bibfnamefont {T.~K.}\ \bibnamefont {Chen}}, \bibinfo {author}
  {\bibfnamefont {T.~W.}\ \bibnamefont {Huang}}, \bibinfo {author}
  {\bibfnamefont {P.~M.}\ \bibnamefont {Wu}}, \bibinfo {author} {\bibfnamefont
  {Y.~C.}\ \bibnamefont {Lee}}, \bibinfo {author} {\bibfnamefont {Y.-L.}\
  \bibnamefont {Huang}}, \bibinfo {author} {\bibfnamefont {Y.-Y.}\ \bibnamefont
  {Chu}}, \bibinfo {author} {\bibfnamefont {D.~C.}\ \bibnamefont {Yan}},\ and\
  \bibinfo {author} {\bibfnamefont {M.~K.}\ \bibnamefont {Wu}},\ }\href
  {https://doi.org/10.1073/pnas.0807325105} {\bibfield  {journal} {\bibinfo
  {journal} {Proc. Nat. Acad. Sci.}\ }\textbf {\bibinfo {volume} {105}},\
  \bibinfo {pages} {14262} (\bibinfo {year} {2008})}\BibitemShut {NoStop}%
\bibitem [{\citenamefont {Sales}\ \emph {et~al.}(2009)\citenamefont {Sales},
  \citenamefont {Sefat}, \citenamefont {McGuire}, \citenamefont {Jin},
  \citenamefont {Mandrus},\ and\ \citenamefont {Mozharivskyj}}]{SalesPRB}%
  \BibitemOpen
  \bibfield  {author} {\bibinfo {author} {\bibfnamefont {B.~C.}\ \bibnamefont
  {Sales}}, \bibinfo {author} {\bibfnamefont {A.~S.}\ \bibnamefont {Sefat}},
  \bibinfo {author} {\bibfnamefont {M.~A.}\ \bibnamefont {McGuire}}, \bibinfo
  {author} {\bibfnamefont {R.~Y.}\ \bibnamefont {Jin}}, \bibinfo {author}
  {\bibfnamefont {D.}~\bibnamefont {Mandrus}},\ and\ \bibinfo {author}
  {\bibfnamefont {Y.}~\bibnamefont {Mozharivskyj}},\ }\href
  {https://doi.org/10.1103/PhysRevB.79.094521} {\bibfield  {journal} {\bibinfo
  {journal} {Phys. Rev. B}\ }\textbf {\bibinfo {volume} {79}},\ \bibinfo
  {pages} {094521} (\bibinfo {year} {2009})}\BibitemShut {NoStop}%
\bibitem [{\citenamefont {Medvedev}\ \emph {et~al.}(2009)\citenamefont
  {Medvedev}, \citenamefont {McQueen}, \citenamefont {Troyan}, \citenamefont
  {Palasyuk}, \citenamefont {Eremets}, \citenamefont {Cava}, \citenamefont
  {Naghavi}, \citenamefont {Casper}, \citenamefont {Ksenofontov}, \citenamefont
  {Wortmann},\ and\ \citenamefont {Felser}}]{MedvedevNatMat}%
  \BibitemOpen
  \bibfield  {author} {\bibinfo {author} {\bibfnamefont {S.}~\bibnamefont
  {Medvedev}}, \bibinfo {author} {\bibfnamefont {T.~M.}\ \bibnamefont
  {McQueen}}, \bibinfo {author} {\bibfnamefont {I.~A.}\ \bibnamefont {Troyan}},
  \bibinfo {author} {\bibfnamefont {T.}~\bibnamefont {Palasyuk}}, \bibinfo
  {author} {\bibfnamefont {M.~I.}\ \bibnamefont {Eremets}}, \bibinfo {author}
  {\bibfnamefont {R.~J.}\ \bibnamefont {Cava}}, \bibinfo {author}
  {\bibfnamefont {S.}~\bibnamefont {Naghavi}}, \bibinfo {author} {\bibfnamefont
  {F.}~\bibnamefont {Casper}}, \bibinfo {author} {\bibfnamefont
  {V.}~\bibnamefont {Ksenofontov}}, \bibinfo {author} {\bibfnamefont
  {G.}~\bibnamefont {Wortmann}},\ and\ \bibinfo {author} {\bibfnamefont
  {C.}~\bibnamefont {Felser}},\ }\href {https://doi.org/10.1038/nmat2491}
  {\bibfield  {journal} {\bibinfo  {journal} {Nat. Mater.}\ }\textbf {\bibinfo
  {volume} {8}},\ \bibinfo {pages} {630} (\bibinfo {year} {2009})}\BibitemShut
  {NoStop}%
\bibitem [{\citenamefont {Burrard-Lucas}\ \emph {et~al.}(2013)\citenamefont
  {Burrard-Lucas}, \citenamefont {Free}, \citenamefont {Sedlmaier},
  \citenamefont {Wright}, \citenamefont {Cassidy}, \citenamefont {Hara},
  \citenamefont {Corkett}, \citenamefont {Lancaster}, \citenamefont {Baker},
  \citenamefont {Blundell},\ and\ \citenamefont {Clarke}}]{BurrardNatMat}%
  \BibitemOpen
  \bibfield  {author} {\bibinfo {author} {\bibfnamefont {M.}~\bibnamefont
  {Burrard-Lucas}}, \bibinfo {author} {\bibfnamefont {D.~G.}\ \bibnamefont
  {Free}}, \bibinfo {author} {\bibfnamefont {S.~J.}\ \bibnamefont {Sedlmaier}},
  \bibinfo {author} {\bibfnamefont {J.~D.}\ \bibnamefont {Wright}}, \bibinfo
  {author} {\bibfnamefont {S.~J.}\ \bibnamefont {Cassidy}}, \bibinfo {author}
  {\bibfnamefont {Y.}~\bibnamefont {Hara}}, \bibinfo {author} {\bibfnamefont
  {A.~J.}\ \bibnamefont {Corkett}}, \bibinfo {author} {\bibfnamefont
  {T.}~\bibnamefont {Lancaster}}, \bibinfo {author} {\bibfnamefont {P.~J.}\
  \bibnamefont {Baker}}, \bibinfo {author} {\bibfnamefont {S.~J.}\ \bibnamefont
  {Blundell}},\ and\ \bibinfo {author} {\bibfnamefont {S.~J.}\ \bibnamefont
  {Clarke}},\ }\href {https://doi.org/10.1038/nmat3464} {\bibfield  {journal}
  {\bibinfo  {journal} {Nat. Mater.}\ }\textbf {\bibinfo {volume} {12}},\
  \bibinfo {pages} {15} (\bibinfo {year} {2013})}\BibitemShut {NoStop}%
\bibitem [{\citenamefont {Meng}\ \emph {et~al.}(2022)\citenamefont {Meng},
  \citenamefont {Xing}, \citenamefont {Yi}, \citenamefont {Li}, \citenamefont
  {Zhou}, \citenamefont {Li}, \citenamefont {Zhang}, \citenamefont {Wei},
  \citenamefont {Feng}, \citenamefont {Terashima}, \citenamefont {Takano},
  \citenamefont {Sun},\ and\ \citenamefont {Shi}}]{MengyanPhysRevB.105.134506}%
  \BibitemOpen
  \bibfield  {author} {\bibinfo {author} {\bibfnamefont {Y.}~\bibnamefont
  {Meng}}, \bibinfo {author} {\bibfnamefont {X.}~\bibnamefont {Xing}}, \bibinfo
  {author} {\bibfnamefont {X.}~\bibnamefont {Yi}}, \bibinfo {author}
  {\bibfnamefont {B.}~\bibnamefont {Li}}, \bibinfo {author} {\bibfnamefont
  {N.}~\bibnamefont {Zhou}}, \bibinfo {author} {\bibfnamefont {M.}~\bibnamefont
  {Li}}, \bibinfo {author} {\bibfnamefont {Y.}~\bibnamefont {Zhang}}, \bibinfo
  {author} {\bibfnamefont {W.}~\bibnamefont {Wei}}, \bibinfo {author}
  {\bibfnamefont {J.}~\bibnamefont {Feng}}, \bibinfo {author} {\bibfnamefont
  {K.}~\bibnamefont {Terashima}}, \bibinfo {author} {\bibfnamefont
  {Y.}~\bibnamefont {Takano}}, \bibinfo {author} {\bibfnamefont
  {Y.}~\bibnamefont {Sun}},\ and\ \bibinfo {author} {\bibfnamefont
  {Z.}~\bibnamefont {Shi}},\ }\href
  {https://doi.org/10.1103/PhysRevB.105.134506} {\bibfield  {journal} {\bibinfo
   {journal} {Phys. Rev. B}\ }\textbf {\bibinfo {volume} {105}},\ \bibinfo
  {pages} {134506} (\bibinfo {year} {2022})}\BibitemShut {NoStop}%
\bibitem [{\citenamefont {Ge}\ \emph {et~al.}(2015)\citenamefont {Ge},
  \citenamefont {Liu}, \citenamefont {Liu}, \citenamefont {Gao}, \citenamefont
  {Qian}, \citenamefont {Zhang},\ and\ \citenamefont {Jia}}]{GeNatMatter}%
  \BibitemOpen
  \bibfield  {author} {\bibinfo {author} {\bibfnamefont {J.~F.}\ \bibnamefont
  {Ge}}, \bibinfo {author} {\bibfnamefont {Z.~L.}\ \bibnamefont {Liu}},
  \bibinfo {author} {\bibfnamefont {C.}~\bibnamefont {Liu}}, \bibinfo {author}
  {\bibfnamefont {C.~L.}\ \bibnamefont {Gao}}, \bibinfo {author} {\bibfnamefont
  {D.}~\bibnamefont {Qian}}, \bibinfo {author} {\bibfnamefont {Q.~M.}\
  \bibnamefont {Zhang}},\ and\ \bibinfo {author} {\bibfnamefont {J.~F.}\
  \bibnamefont {Jia}},\ }\href {https://doi.org/10.1038/nmat4153} {\bibfield
  {journal} {\bibinfo  {journal} {Nature Materials}\ }\textbf {\bibinfo
  {volume} {14}},\ \bibinfo {pages} {285} (\bibinfo {year} {2015})}\BibitemShut
  {NoStop}%
\bibitem [{\citenamefont {Sun}\ \emph {et~al.}(2016)\citenamefont {Sun},
  \citenamefont {Pyon},\ and\ \citenamefont {Tamegai}}]{SunPhysRevB.93.104502}%
  \BibitemOpen
  \bibfield  {author} {\bibinfo {author} {\bibfnamefont {Y.}~\bibnamefont
  {Sun}}, \bibinfo {author} {\bibfnamefont {S.}~\bibnamefont {Pyon}},\ and\
  \bibinfo {author} {\bibfnamefont {T.}~\bibnamefont {Tamegai}},\ }\href
  {https://doi.org/10.1103/PhysRevB.93.104502} {\bibfield  {journal} {\bibinfo
  {journal} {Phys. Rev. B}\ }\textbf {\bibinfo {volume} {93}},\ \bibinfo
  {pages} {104502} (\bibinfo {year} {2016})}\BibitemShut {NoStop}%
\bibitem [{\citenamefont {Sun}\ \emph {et~al.}(2019)\citenamefont {Sun},
  \citenamefont {Shi},\ and\ \citenamefont {Tamegai}}]{Sun_sustreview}%
  \BibitemOpen
  \bibfield  {author} {\bibinfo {author} {\bibfnamefont {Y.}~\bibnamefont
  {Sun}}, \bibinfo {author} {\bibfnamefont {Z.}~\bibnamefont {Shi}},\ and\
  \bibinfo {author} {\bibfnamefont {T.}~\bibnamefont {Tamegai}},\ }\href
  {https://doi.org/10.1088/1361-6668/ab30c2} {\bibfield  {journal} {\bibinfo
  {journal} {Supercond. Sci. Technol.}\ }\textbf {\bibinfo {volume} {32}},\
  \bibinfo {pages} {103001} (\bibinfo {year} {2019})}\BibitemShut {NoStop}%
\bibitem [{\citenamefont {Hou}\ \emph {et~al.}(2024)\citenamefont {Hou},
  \citenamefont {Wei}, \citenamefont {Zhou}, \citenamefont {Liu}, \citenamefont
  {Wang}, \citenamefont {Xing}, \citenamefont {Zhang}, \citenamefont {Zhou},
  \citenamefont {Pan}, \citenamefont {Sun},\ and\ \citenamefont
  {Shi}}]{HouPNAS}%
  \BibitemOpen
  \bibfield  {author} {\bibinfo {author} {\bibfnamefont {Q.}~\bibnamefont
  {Hou}}, \bibinfo {author} {\bibfnamefont {W.}~\bibnamefont {Wei}}, \bibinfo
  {author} {\bibfnamefont {X.}~\bibnamefont {Zhou}}, \bibinfo {author}
  {\bibfnamefont {W.}~\bibnamefont {Liu}}, \bibinfo {author} {\bibfnamefont
  {K.}~\bibnamefont {Wang}}, \bibinfo {author} {\bibfnamefont {X.}~\bibnamefont
  {Xing}}, \bibinfo {author} {\bibfnamefont {Y.}~\bibnamefont {Zhang}},
  \bibinfo {author} {\bibfnamefont {N.}~\bibnamefont {Zhou}}, \bibinfo {author}
  {\bibfnamefont {Y.}~\bibnamefont {Pan}}, \bibinfo {author} {\bibfnamefont
  {Y.}~\bibnamefont {Sun}},\ and\ \bibinfo {author} {\bibfnamefont
  {Z.}~\bibnamefont {Shi}},\ }\href {https://doi.org/10.1073/pnas.2409756121}
  {\bibfield  {journal} {\bibinfo  {journal} {Proceedings of the National
  Academy of Sciences}\ }\textbf {\bibinfo {volume} {121}},\ \bibinfo {pages}
  {e2409756121} (\bibinfo {year} {2024})}\BibitemShut {NoStop}%
\bibitem [{\citenamefont {Moll}(2018)}]{Mollreview}%
  \BibitemOpen
  \bibfield  {author} {\bibinfo {author} {\bibfnamefont {P.~J.}\ \bibnamefont
  {Moll}},\ }\href {https://doi.org/10.1146/annurev-conmatphys-033117-054021}
  {\bibfield  {journal} {\bibinfo  {journal} {Annu. Rev. Condens. Matter
  Phys.}\ }\textbf {\bibinfo {volume} {9}},\ \bibinfo {pages} {147} (\bibinfo
  {year} {2018})}\BibitemShut {NoStop}%
\bibitem [{\citenamefont {Miyazawa}\ \emph {et~al.}(2021)\citenamefont
  {Miyazawa}, \citenamefont {Tadokoro}, \citenamefont {Horikawa}, \citenamefont
  {Tamegai}, \citenamefont {Sun},\ and\ \citenamefont
  {Kitano}}]{Miyazawa_2021}%
  \BibitemOpen
  \bibfield  {author} {\bibinfo {author} {\bibfnamefont {T.}~\bibnamefont
  {Miyazawa}}, \bibinfo {author} {\bibfnamefont {N.}~\bibnamefont {Tadokoro}},
  \bibinfo {author} {\bibfnamefont {S.}~\bibnamefont {Horikawa}}, \bibinfo
  {author} {\bibfnamefont {T.}~\bibnamefont {Tamegai}}, \bibinfo {author}
  {\bibfnamefont {Y.}~\bibnamefont {Sun}},\ and\ \bibinfo {author}
  {\bibfnamefont {H.}~\bibnamefont {Kitano}},\ }\href
  {https://doi.org/10.1088/1742-6596/1975/1/012010} {\bibfield  {journal}
  {\bibinfo  {journal} {Journal of Physics: Conference Series}\ }\textbf
  {\bibinfo {volume} {1975}},\ \bibinfo {pages} {012010} (\bibinfo {year}
  {2021})}\BibitemShut {NoStop}%
\bibitem [{sup()}]{supplement}%
  \BibitemOpen
  \href@noop {} {\ }\bibinfo {note} {{See Supplemental Material at [] for the
  MHLs and the estimated $J_{\rm{c}}$ based on the Bean model.}}\BibitemShut
  {Stop}%
\bibitem [{\citenamefont {Sun}\ \emph {et~al.}(2015)\citenamefont {Sun},
  \citenamefont {Pyon}, \citenamefont {Tamegai}, \citenamefont {Kobayashi},
  \citenamefont {Watashige}, \citenamefont {Kasahara}, \citenamefont
  {Matsuda},\ and\ \citenamefont {Shibauchi}}]{SunPhysRevB.92.144509}%
  \BibitemOpen
  \bibfield  {author} {\bibinfo {author} {\bibfnamefont {Y.}~\bibnamefont
  {Sun}}, \bibinfo {author} {\bibfnamefont {S.}~\bibnamefont {Pyon}}, \bibinfo
  {author} {\bibfnamefont {T.}~\bibnamefont {Tamegai}}, \bibinfo {author}
  {\bibfnamefont {R.}~\bibnamefont {Kobayashi}}, \bibinfo {author}
  {\bibfnamefont {T.}~\bibnamefont {Watashige}}, \bibinfo {author}
  {\bibfnamefont {S.}~\bibnamefont {Kasahara}}, \bibinfo {author}
  {\bibfnamefont {Y.}~\bibnamefont {Matsuda}},\ and\ \bibinfo {author}
  {\bibfnamefont {T.}~\bibnamefont {Shibauchi}},\ }\href
  {https://doi.org/10.1103/PhysRevB.92.144509} {\bibfield  {journal} {\bibinfo
  {journal} {Phys. Rev. B}\ }\textbf {\bibinfo {volume} {92}},\ \bibinfo
  {pages} {144509} (\bibinfo {year} {2015})}\BibitemShut {NoStop}%
\bibitem [{\citenamefont {Prozorov}\ \emph {et~al.}(2008)\citenamefont
  {Prozorov}, \citenamefont {Ni}, \citenamefont {Tanatar}, \citenamefont
  {Kogan}, \citenamefont {Gordon}, \citenamefont {Martin}, \citenamefont
  {Blomberg}, \citenamefont {Prommapan}, \citenamefont {Yan}, \citenamefont
  {Bud'ko},\ and\ \citenamefont {Canfield}}]{ProzorovPRBBaCoJc}%
  \BibitemOpen
  \bibfield  {author} {\bibinfo {author} {\bibfnamefont {R.}~\bibnamefont
  {Prozorov}}, \bibinfo {author} {\bibfnamefont {N.}~\bibnamefont {Ni}},
  \bibinfo {author} {\bibfnamefont {M.~A.}\ \bibnamefont {Tanatar}}, \bibinfo
  {author} {\bibfnamefont {V.~G.}\ \bibnamefont {Kogan}}, \bibinfo {author}
  {\bibfnamefont {R.~T.}\ \bibnamefont {Gordon}}, \bibinfo {author}
  {\bibfnamefont {C.}~\bibnamefont {Martin}}, \bibinfo {author} {\bibfnamefont
  {E.~C.}\ \bibnamefont {Blomberg}}, \bibinfo {author} {\bibfnamefont
  {P.}~\bibnamefont {Prommapan}}, \bibinfo {author} {\bibfnamefont {J.~Q.}\
  \bibnamefont {Yan}}, \bibinfo {author} {\bibfnamefont {S.~L.}\ \bibnamefont
  {Bud'ko}},\ and\ \bibinfo {author} {\bibfnamefont {P.~C.}\ \bibnamefont
  {Canfield}},\ }\href {https://doi.org/10.1103/PhysRevB.78.224506} {\bibfield
  {journal} {\bibinfo  {journal} {Phys. Rev. B}\ }\textbf {\bibinfo {volume}
  {78}},\ \bibinfo {pages} {224506} (\bibinfo {year} {2008})}\BibitemShut
  {NoStop}%
\bibitem [{\citenamefont {Kupriyanov}\ and\ \citenamefont
  {Lukichev}(1980)}]{KLtheory}%
  \BibitemOpen
  \bibfield  {author} {\bibinfo {author} {\bibfnamefont {M.~Y.}\ \bibnamefont
  {Kupriyanov}}\ and\ \bibinfo {author} {\bibfnamefont {V.~F.}\ \bibnamefont
  {Lukichev}},\ }\href@noop {} {\bibfield  {journal} {\bibinfo  {journal} {Fiz.
  Nizk. Temp.}\ }\textbf {\bibinfo {volume} {6}},\ \bibinfo {pages} {445}
  (\bibinfo {year} {1980})},\ \bibinfo {note} {[Sov. J. Low Temp. Phys. 6, 210
  (1980)]}\BibitemShut {NoStop}%
\bibitem [{\citenamefont {Kasahara}\ \emph {et~al.}(2014)\citenamefont
  {Kasahara}, \citenamefont {Watashige}, \citenamefont {Hanaguri},
  \citenamefont {Kohsaka}, \citenamefont {Yamashita}, \citenamefont
  {Shimoyama}, \citenamefont {Mizukami}, \citenamefont {Endo}, \citenamefont
  {Ikeda}, \citenamefont {Aoyama}, \citenamefont {Terashima}, \citenamefont
  {Uji}, \citenamefont {Wolf}, \citenamefont {von Löhneysen}, \citenamefont
  {Shibauchi},\ and\ \citenamefont {Matsuda}}]{Kasahara18112014}%
  \BibitemOpen
  \bibfield  {author} {\bibinfo {author} {\bibfnamefont {S.}~\bibnamefont
  {Kasahara}}, \bibinfo {author} {\bibfnamefont {T.}~\bibnamefont {Watashige}},
  \bibinfo {author} {\bibfnamefont {T.}~\bibnamefont {Hanaguri}}, \bibinfo
  {author} {\bibfnamefont {Y.}~\bibnamefont {Kohsaka}}, \bibinfo {author}
  {\bibfnamefont {T.}~\bibnamefont {Yamashita}}, \bibinfo {author}
  {\bibfnamefont {Y.}~\bibnamefont {Shimoyama}}, \bibinfo {author}
  {\bibfnamefont {Y.}~\bibnamefont {Mizukami}}, \bibinfo {author}
  {\bibfnamefont {R.}~\bibnamefont {Endo}}, \bibinfo {author} {\bibfnamefont
  {H.}~\bibnamefont {Ikeda}}, \bibinfo {author} {\bibfnamefont
  {K.}~\bibnamefont {Aoyama}}, \bibinfo {author} {\bibfnamefont
  {T.}~\bibnamefont {Terashima}}, \bibinfo {author} {\bibfnamefont
  {S.}~\bibnamefont {Uji}}, \bibinfo {author} {\bibfnamefont {T.}~\bibnamefont
  {Wolf}}, \bibinfo {author} {\bibfnamefont {H.}~\bibnamefont {von
  Löhneysen}}, \bibinfo {author} {\bibfnamefont {T.}~\bibnamefont
  {Shibauchi}},\ and\ \bibinfo {author} {\bibfnamefont {Y.}~\bibnamefont
  {Matsuda}},\ }\bibfield  {title} {\bibinfo {title} {Field-induced
  superconducting phase of fese in the bcs-bec cross-over},\ }\href
  {https://doi.org/10.1073/pnas.1413477111} {\bibfield  {journal} {\bibinfo
  {journal} {Proc. Nat. Acad. Sci.}\ }\textbf {\bibinfo {volume} {111}},\
  \bibinfo {pages} {16309} (\bibinfo {year} {2014})}\BibitemShut {NoStop}%
\bibitem [{\citenamefont {Shibauchi}\ \emph {et~al.}(2020)\citenamefont
  {Shibauchi}, \citenamefont {Hanaguri},\ and\ \citenamefont
  {Matsuda}}]{ShibauchiFeSereview}%
  \BibitemOpen
  \bibfield  {author} {\bibinfo {author} {\bibfnamefont {T.}~\bibnamefont
  {Shibauchi}}, \bibinfo {author} {\bibfnamefont {T.}~\bibnamefont
  {Hanaguri}},\ and\ \bibinfo {author} {\bibfnamefont {Y.}~\bibnamefont
  {Matsuda}},\ }\href {https://doi.org/10.7566/JPSJ.89.102002} {\bibfield
  {journal} {\bibinfo  {journal} {Journal of the Physical Society of Japan}\
  }\textbf {\bibinfo {volume} {89}},\ \bibinfo {pages} {102002} (\bibinfo
  {year} {2020})}\BibitemShut {NoStop}%
\bibitem [{\citenamefont {Nider\"{o}st}\ \emph {et~al.}(1998)\citenamefont
  {Nider\"{o}st}, \citenamefont {Frassanito}, \citenamefont {Saalfrank},
  \citenamefont {Mota}, \citenamefont {Blatter}, \citenamefont {Zavaritsky},
  \citenamefont {Li},\ and\ \citenamefont {Kes}}]{Nideroest1998}%
  \BibitemOpen
  \bibfield  {author} {\bibinfo {author} {\bibfnamefont {M.}~\bibnamefont
  {Nider\"{o}st}}, \bibinfo {author} {\bibfnamefont {R.}~\bibnamefont
  {Frassanito}}, \bibinfo {author} {\bibfnamefont {M.}~\bibnamefont
  {Saalfrank}}, \bibinfo {author} {\bibfnamefont {A.~C.}\ \bibnamefont {Mota}},
  \bibinfo {author} {\bibfnamefont {G.}~\bibnamefont {Blatter}}, \bibinfo
  {author} {\bibfnamefont {V.~N.}\ \bibnamefont {Zavaritsky}}, \bibinfo
  {author} {\bibfnamefont {T.~W.}\ \bibnamefont {Li}},\ and\ \bibinfo {author}
  {\bibfnamefont {P.~H.}\ \bibnamefont {Kes}},\ }\href
  {https://doi.org/10.1103/PhysRevLett.81.3231} {\bibfield  {journal} {\bibinfo
   {journal} {Phys. Rev. Lett.}\ }\textbf {\bibinfo {volume} {81}},\ \bibinfo
  {pages} {3231} (\bibinfo {year} {1998})}\BibitemShut {NoStop}%
\bibitem [{\citenamefont {Joshi}\ \emph {et~al.}(2019)\citenamefont {Joshi},
  \citenamefont {Nusran}, \citenamefont {Tanatar}, \citenamefont {Cho},
  \citenamefont {Meier}, \citenamefont {Bud'ko}, \citenamefont {Canfield},\
  and\ \citenamefont {Prozorov}}]{JoshiPhysRevApplied.11.014035}%
  \BibitemOpen
  \bibfield  {author} {\bibinfo {author} {\bibfnamefont {K.}~\bibnamefont
  {Joshi}}, \bibinfo {author} {\bibfnamefont {N.}~\bibnamefont {Nusran}},
  \bibinfo {author} {\bibfnamefont {M.}~\bibnamefont {Tanatar}}, \bibinfo
  {author} {\bibfnamefont {K.}~\bibnamefont {Cho}}, \bibinfo {author}
  {\bibfnamefont {W.}~\bibnamefont {Meier}}, \bibinfo {author} {\bibfnamefont
  {S.}~\bibnamefont {Bud'ko}}, \bibinfo {author} {\bibfnamefont
  {P.}~\bibnamefont {Canfield}},\ and\ \bibinfo {author} {\bibfnamefont
  {R.}~\bibnamefont {Prozorov}},\ }\href
  {https://doi.org/10.1103/PhysRevApplied.11.014035} {\bibfield  {journal}
  {\bibinfo  {journal} {Phys. Rev. Appl.}\ }\textbf {\bibinfo {volume} {11}},\
  \bibinfo {pages} {014035} (\bibinfo {year} {2019})}\BibitemShut {NoStop}%
\bibitem [{\citenamefont {Song}\ \emph {et~al.}(2011)\citenamefont {Song},
  \citenamefont {Ghim}, \citenamefont {Yoon}, \citenamefont {Lee},
  \citenamefont {Jung}, \citenamefont {Ji}, \citenamefont {Shim}, \citenamefont
  {Bang},\ and\ \citenamefont {Kwon}}]{Song_2011LiFeSepenetrationdepth}%
  \BibitemOpen
  \bibfield  {author} {\bibinfo {author} {\bibfnamefont {Y.~J.}\ \bibnamefont
  {Song}}, \bibinfo {author} {\bibfnamefont {J.~S.}\ \bibnamefont {Ghim}},
  \bibinfo {author} {\bibfnamefont {J.~H.}\ \bibnamefont {Yoon}}, \bibinfo
  {author} {\bibfnamefont {K.~J.}\ \bibnamefont {Lee}}, \bibinfo {author}
  {\bibfnamefont {M.~H.}\ \bibnamefont {Jung}}, \bibinfo {author}
  {\bibfnamefont {H.-S.}\ \bibnamefont {Ji}}, \bibinfo {author} {\bibfnamefont
  {J.~H.}\ \bibnamefont {Shim}}, \bibinfo {author} {\bibfnamefont
  {Y.}~\bibnamefont {Bang}},\ and\ \bibinfo {author} {\bibfnamefont {Y.~S.}\
  \bibnamefont {Kwon}},\ }\href {https://doi.org/10.1209/0295-5075/94/57008}
  {\bibfield  {journal} {\bibinfo  {journal} {EPL}\ }\textbf {\bibinfo {volume}
  {94}},\ \bibinfo {pages} {57008} (\bibinfo {year} {2011})}\BibitemShut
  {NoStop}%
\bibitem [{\citenamefont {Abdel-Hafiez}\ \emph {et~al.}(2013)\citenamefont
  {Abdel-Hafiez}, \citenamefont {Ge}, \citenamefont {Vasiliev}, \citenamefont
  {Chareev}, \citenamefont {Van~de Vondel}, \citenamefont {Moshchalkov},\ and\
  \citenamefont {Silhanek}}]{AbdelFeSeHc1PhysRevB.88.174512}%
  \BibitemOpen
  \bibfield  {author} {\bibinfo {author} {\bibfnamefont {M.}~\bibnamefont
  {Abdel-Hafiez}}, \bibinfo {author} {\bibfnamefont {J.}~\bibnamefont {Ge}},
  \bibinfo {author} {\bibfnamefont {A.~N.}\ \bibnamefont {Vasiliev}}, \bibinfo
  {author} {\bibfnamefont {D.~A.}\ \bibnamefont {Chareev}}, \bibinfo {author}
  {\bibfnamefont {J.}~\bibnamefont {Van~de Vondel}}, \bibinfo {author}
  {\bibfnamefont {V.~V.}\ \bibnamefont {Moshchalkov}},\ and\ \bibinfo {author}
  {\bibfnamefont {A.~V.}\ \bibnamefont {Silhanek}},\ }\href
  {https://doi.org/10.1103/PhysRevB.88.174512} {\bibfield  {journal} {\bibinfo
  {journal} {Phys. Rev. B}\ }\textbf {\bibinfo {volume} {88}},\ \bibinfo
  {pages} {174512} (\bibinfo {year} {2013})}\BibitemShut {NoStop}%
\bibitem [{\citenamefont {Abrikosov}(1964)}]{AbrikosovHc1}%
  \BibitemOpen
  \bibfield  {author} {\bibinfo {author} {\bibfnamefont {A.~A.}\ \bibnamefont
  {Abrikosov}},\ }\href@noop {} {\bibfield  {journal} {\bibinfo  {journal}
  {Soviet Physics JETP}\ }\textbf {\bibinfo {volume} {19}},\ \bibinfo {pages}
  {988} (\bibinfo {year} {1964})}\BibitemShut {NoStop}%
\bibitem [{\citenamefont {Galluzzi}\ \emph {et~al.}(2023)\citenamefont
  {Galluzzi}, \citenamefont {Buchkov}, \citenamefont {Tomov}, \citenamefont
  {Nazarova}, \citenamefont {Leo}, \citenamefont {Grimaldi}, \citenamefont
  {Crisan},\ and\ \citenamefont {Polichetti}}]{Galluzzicondmat8040091}%
  \BibitemOpen
  \bibfield  {author} {\bibinfo {author} {\bibfnamefont {A.}~\bibnamefont
  {Galluzzi}}, \bibinfo {author} {\bibfnamefont {K.}~\bibnamefont {Buchkov}},
  \bibinfo {author} {\bibfnamefont {V.}~\bibnamefont {Tomov}}, \bibinfo
  {author} {\bibfnamefont {E.}~\bibnamefont {Nazarova}}, \bibinfo {author}
  {\bibfnamefont {A.}~\bibnamefont {Leo}}, \bibinfo {author} {\bibfnamefont
  {G.}~\bibnamefont {Grimaldi}}, \bibinfo {author} {\bibfnamefont
  {A.}~\bibnamefont {Crisan}},\ and\ \bibinfo {author} {\bibfnamefont
  {M.}~\bibnamefont {Polichetti}},\ }\href
  {https://doi.org/10.3390/condmat8040091} {\bibfield  {journal} {\bibinfo
  {journal} {Condensed Matter}\ }\textbf {\bibinfo {volume} {8}},\ \bibinfo
  {pages} {91} (\bibinfo {year} {2023})}\BibitemShut {NoStop}%
\end{thebibliography}%

\end{document}